\documentclass[aps,preprint,floatfix,nofootinbib,showpacs]{revtex4}
\usepackage{graphicx,color,bm}
\usepackage{hyperref}
\usepackage{dcolumn}
\usepackage{amssymb}
\usepackage{amsmath}
\usepackage{epsfig}    
\usepackage{slashed}

\def\be{\begin{equation}}
\def\ee{\end{equation}}
\newcommand{\bea}{\begin{eqnarray}}
\newcommand{\eea}{\end{eqnarray}}
\newcommand{\nn}{\nonumber}

\begin{document}
\title{Accommodation of the Dirac Phase in \\
  the Krauss-Nasri-Trodden Model}
\author{
Kingman Cheung$^{1,2,3}$, Hiroyuki Ishida$^1$, and Hiroshi Okada$^1$}
\affiliation{
$^1$ Physics Division, National Center for Theoretical Sciences, Hsinchu 30013,
Taiwan\\
$^2$ Department of Physics, National Tsing Hua University,
Hsinchu 300, Taiwan \\
$^3$ Division of Quantum Phases and Devices, School of Physics, 
Konkuk University, Seoul 143-701, Republic of Korea \\
}
\date{\today}

\begin{abstract}
We investigate one of the radiative models, Kraus-Nasri-Trodden model ,
with the maximal value of Dirac CP violating phase, $\delta$ is $-\pi/2$ (or equivalently $3\pi/2$), 
which is preferred in not only recent long baseline experiments but also the global fit of neutrino oscillation data. 
We show that our predicted region of the $\mu \mathchar`- e$ conversion rate 
can be searched in the future experiments without conflicting lepton-flavor violation and dark matter constraints. 
\end{abstract}

\maketitle

\section{Introduction}

We have observed more matter than antimatter in our daily lives, e.g.,
more protons than antiprotons, more electrons than positrons, and more
hydrogen than anti-hydrogen. The list can go on and on.  Indeed,
experimentalists have also observed more matter than anti-matter in
cosmic ray experiments. Such an asymmetry is known as matter-antimatter
asymmetry.
Charge-Parity ($CP$) violation is one of the key ingredients to the 
understanding of the evolution in the early Universe why we have 
observed more matter than antimatter nowadays. 
$CP$ violation was first observed in the Kaon system in early 60's
\cite{kaon}. It was only {evident} until early 2000 that $CP$ violation was
observed in the $B$-meson system \cite{bmeson}.  Both Kaon and $B$-meson
$CP$ violation data can be accommodated by the so-called 
Kobayashi-Maskawa (KM) mixing
matrix in the quark sector \cite{km} within the standard model (SM).
It is well-known that the amount of $CP$ violation allowed by the SM
is not large enough to explain the matter-antimatter asymmetry of the
Universe.  Further sources of $CP$ violations are hot topics for
physics beyond the SM.

Recently, the T2K experiment reported measurements of appearance rates for
$\nu_\mu \to \nu_e$ and $\bar\nu_\mu \to \bar \nu_e$, and they 
found that they indeed have different rates.  Thus, it is a hint of 
$CP$ violation and thus resulting in nonzero values for the $CP$-odd
phase $\delta$.  The data preferred maximal $\theta_{23}$ mixing, 
$\delta_{CP} \sim - \pi/2$ (or equivalently $3\pi/2$), 
and normal mass hierarchy (NH) 
over the inverted mass hierarchy (IH) \cite{t2k-talk}.
The fitted range for $\delta_{CP}$ is given by
\begin{equation}
 \delta_{CP} = \left[  - 3.13, - 0.39  \right] \;\;({\rm NH}), \qquad
 \delta_{CP} = \left[  - 2.09, - 0.74  \right] \;\; ({\rm IH}) 
\end{equation}
at 90\% CL, with the best fit at around $\delta_{CP}\sim -\pi/2 \;({\rm NH})$ 
(or equivalently $3\pi/2$).
Also, the experiment claimed a 90\%CL exclusion of $\delta = 0$ and $\pi$.
This is consistent with the most recent global analysis of neutrino
oscillation data \cite{global}.

In this work, we show that the radiative neutrino-mass model,
due to Krauss, Nasri, and Trodden (KNT) \cite{knt}, can accommodate the $CP$-odd
phase with the choice of the complex $f_{\alpha\beta}$ parameters.
The KNT model generates tiny neutrino masses based on a 3-loop diagram
with right-handed (RH) neutrinos at TeV scale and a $Z_2$ symmetry to avoid
the type-I see-saw mass. It was also shown that the TeV scale RH neutrinos can be
the dark matter candidate and searchable at the future linear colliders
\cite{seto}.  

We shall extend the model by employing three RH neutrinos and complex
$f_{\alpha\beta}$ parameters, so that we can satisfy not only neutrino
oscillation data and dark matter constraints, but also the
lepton-flavor violations, as well as favors a nonzero $CP$-odd
phase. The whole setup is consistent with neutrino oscillation data, 
lepton-flavor
violations, $\mu$-$e$ conversion, and dark matter constraints.

The paper is organized as follows. 
In the next section, we describe the KNT model with 3 RH neutrinos
and the neutrino mass matrix, as well as 
the constraints and phenomenology of the model,
such as lepton-flavor violations, dark matter, and collider physics.
We also show numerically that the model is consistent with all the data.
Section~III is devoted for conclusions and discussion.

\section{Model}
In this section, we briefly describe the KNT model and the corresponding
active neutrino mass matrix, as well as all the existing constraints.

 \begin{widetext}
\begin{center} 
\begin{table}
\begin{tabular}{|c||c|c|c||c|c|c|}\hline\hline  
& ~ $L_{L_i}$ ~ & ~ $e_{R_i}$ ~ & ~ $N_{R_i}$ ~ &~ $\Phi$ ~ & ~ $S^{+}_{1}$ ~ & ~ $S^{+}_{2}$ ~
\\\hline 
$SU(2)_L$ & $\bm{2}$  & $\bm{1}$  & $\bm{1}$  & $\bm{2}$& $\bm{1}$& $\bm{1}$   \\\hline 
$U(1)_Y$ & $-\frac12$ & $-1$  & $0$  & $\frac12$  & $1$  & $1$ \\\hline
$Z_2$ & $+$ & $+$  & $-$ & $+$ & $+$ & $-$  \\\hline
\end{tabular}
\caption{Field contents of the KNT model and their charge assignments 
under $SU(2)_L\times U(1)_Y\times Z_2$, where the lower index 
$i(=1\mathchar`-3)$ represents the generation. }
\label{tab:1}
\end{table}
\end{center}
\end{widetext}
\subsection{Model setup}
We show all the field contents and their charge assignments 
in Table~\ref{tab:1}.
The relevant Lagrangian and the Higgs potential are, respectively, given by
\begin{align}
\label{eq:Yukawa}
-{\cal L}^{}&=
(y_\ell)_{\alpha}\bar L_{L_\alpha} \Phi e_{R_\beta} + f_{\alpha \beta}\bar L^c_{L_\alpha}(i\sigma_2) L_{L_\beta} S_{1}^{+}
+g_{i \alpha} \bar N^{c}_{R_i} e_{R_\alpha} S_{2}^{+} + M_{N_i} \bar N_{R_i}^c N_{R_i}+{\rm h.c.} ,\\
{\cal V}&=m_\Phi^2 \Phi^\dag \Phi + m_{S_1}^2 S_{1}^{+} S_{1}^{-}
+ m_{S_{2}}^2 S_{2}^{+} S_{2}^{-} + \lambda_0\left[( S_{1}^{+} S_{2}^-)^2+ ( S_{2}^{+} S_{1}^-)^2\right]  
+\lambda_{S_1S_2} (S_{1}^{+} S_{1}^{-})(S_{2}^{+} S_{2}^{-})
\nn\\
&
+\lambda_{S_1} |S_{1}^{+} S_{1}^{-}|^2
+\lambda_{S_2} |S_{2}^{+} S_{2}^{-}|^2 
 +\lambda_{\Phi} |\Phi^\dag \Phi|^2
+\lambda_{\Phi S_{1}} (\Phi^\dag \Phi)(S_{1}^{+} S_{1}^{-})
+\lambda_{\Phi S_{2}} (\Phi^\dag \Phi)(S_{2}^{+} S_{2}^{-}),
\end{align}
where $i,j=1\mathchar`-3$ and $\alpha, \beta = e, \mu, \tau$
are the generation indices, $\sigma_2$ is the second 
component of the Pauli matrices, $f$ is
an anti-symmetric matrix, and we assume $\lambda_0$ to be real for
simplicity. Notice here that the first term in ${\cal L}$ 
induces the charged-lepton
mass eigenstates, (which are symbolized by $m_{\ell_\alpha} \equiv
[m_e,m_\mu,m_\tau]^T$), therefore, the MNS mixing matrix arises from
the neutrino mass matrix only.

{\it Vacuum stability}: Since we have two singly-charged 
scalar bosons, the
pure couplings $\lambda_{S_{1}}$ and $\lambda_{S_2}$ should be greater 
than zero in order to avoid giving them nonzero vacuum 
expectation value (VEV). Therefore,  we have to satisfy the 
following conditions up to the one-loop level:
\begin{align}
&0\lesssim \lambda_{S_1}^{\rm one-loop}\lesssim 4\pi,\quad
  0\lesssim \lambda_{S_2}^{\rm one-loop}\lesssim 4\pi,
\end{align}
with
\begin{align}
&\lambda_{S_1}^{\rm one-loop}=  \lambda_{S_1}-\frac{\lambda_{\Phi S_1}^4 v^4}{3(4\pi)^2m_h^4}
-\frac{\lambda_{\Phi S_2}^2 \lambda_0^2 v^4}{6\pi^2m_{S_2}^4}>0,\\
&\lambda_{S_2}^{\rm one-loop}=  \lambda_{S_2}-\frac{\lambda_{\Phi S_2}^4 v^4}{3(4\pi)^2m_h^4}
-\frac{\lambda_{\Phi S_1}^2 \lambda_0^2 v^4}{6\pi^2m_{S_1}^4}+\frac{4}{(4\pi)^2}\sum_{i,j=1}^3\sum_{\alpha,\beta=1}^3
(g_{i\alpha}M_{N_i} g_{i\beta})(g^*_{j\beta} M_{N_j} g^*_{j\alpha})\nn\\
&\times \int [dx_i]
\frac{\delta(1-x_1-x_2-x_3-x_4)
(1+\delta_{ij}+\delta_{\alpha\beta}+\delta_{ij}\delta_{\alpha\beta}/2)}
{x_1 M^2_{N_i}+x_2 m^2_{\ell_\beta}+x_3 M^2_{N_j}+x_4 m^2_{\ell_\alpha} }
>0,
\end{align}
where $[dx_i]\equiv \Pi_i^4dx_i$, $h$ is the SM Higgs boson, 
$v\approx246$ GeV is VEV of the SM Higgs field, and each of 
$m_{S_1}$ and $m_{S_2}$ is the mass eigenvalue of $S_1^{\pm}$ and $S_2^{\pm}$.
Note that the boson loop gives negative contributions to the quartic 
coupling while the fermion loop gives  positive contributions.

\begin{figure}[t]
\begin{center}
\includegraphics[width=100mm]{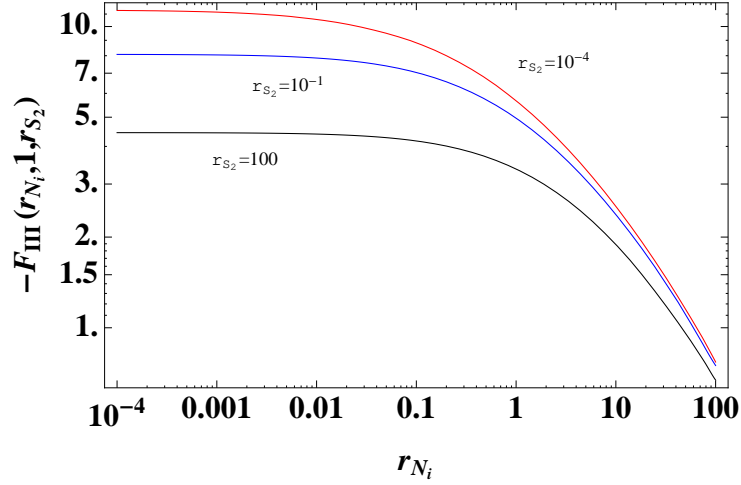}
\caption{Behavior of the loop function $-F_{III}$ versus $r_{N_i}$, 
where we take $M_{\rm max}\equiv m_{S_1}$.
The red line is fixed at $r_{S_2}=10^{-4}$, the blue one at $r_{S_2}=0.1$, 
and the black one at $r_{S_2}=100$.
}
\label{fig:f3}
\end{center}
\end{figure}

\subsection{Active neutrino mass matrix}
The neutrino mass matrix is induced at the three-loop level, and its formula is given by 
\begin{align}
&{\cal M}_{\nu_{ab}}\approx -\frac{4\lambda_0}{(4\pi)^6 M_{\rm Max}^2} 
f_{a\alpha} m_{\ell_\alpha} g^\dag_{\alpha i} M_{N_i} g^*_{i\beta}m_{\ell_\beta} f_{\beta b}
 F_{III}(r_{N_i}, r_{S_1}, r_{S_2}) ,\label{Eq:act_mass}\\
&F_{III}(r_{N_i}, r_{S_1}, r_{S_2})=\int [dx]  \int [dx' ]  \int [dx'']  \nn\\
&\times
\frac{\delta(1-x-y-z)\delta(1-x'-y'-z')\delta(1-x''-y''-z'')}
{x''(z'^2-z') (y r_{S_1}+z r_{S_2}) +y'' (z^2-z) (y' r_{S_1}+z' r_{S_2}) -z'' (z^2-z)(z'^2-z') r_{N_i} },
\end{align}
where $a, b = e, \mu, \tau$, $M_{\rm Max}\equiv$ Max[$M_{N_i},m_{S_1}, m_{S_2}$], 
$r_f\equiv m_f^2/M_{\rm Max}^2$, 
$[dx]\equiv dxdydz$, and we assume that $m_{\ell_\alpha} \ll M_{N_i},m_{S_1}, m_{S_2} $. 
Note here that the three-loop function $F_{III}$ is obtained by 
numerical integration.
Thus, we are preparing an interpolation function to evaluate $F_{III}$ in
our numerical analysis. We show the typical behavior of this function 
in Fig.~\ref{fig:f3}.
Assuming that the mass matrix for the charged-leptons is diagonal,
the neutrino mass matrix ${\cal M}_{\nu_{ab}}$ is diagonalized 
by the MNS mixing matrix $V_{\rm MNS}$.

{\it The normal ordering case}; $ D_\nu\equiv (0,m_{\nu_2},m_{\nu_3})$, is 
written in terms of experimental values as
follows:
\begin{align}
|{\cal M}_{\nu}|
&=|(V_{\rm MNS} D_\nu V_{\rm MNS}^T)|
\nn\\
&\approx \left[\begin{array}{ccc} 
0.0845-0.475 & 0.0629-0.971 &0.0411-0.964 \\
* & 1.44-3.49 &  1.94-2.85 \\
* & * &   1.22-3.33\\
  \end{array}\right]\times 10^{-11}\ {\rm GeV},\label{eq:exp-Neutmass-NH}
\\
V_{\rm MNS}&=
\left[\begin{array}{ccc} {c_{13}}c_{12} &c_{13}s_{12} & s_{13} e^{-i\delta}\\
 -c_{23}s_{12}-s_{23}s_{13}c_{12}e^{i\delta} & c_{23}c_{12}-s_{23}s_{13}s_{12}e^{i\delta} & s_{23}c_{13}\\
  s_{23}s_{12}-c_{23}s_{13}c_{12}e^{i\delta} & -s_{23}c_{12}-c_{23}s_{13}s_{12}e^{i\delta} & c_{23}c_{13}\\
  \end{array}\right]
\left[\begin{array}{ccc} e^{i\alpha_1/2} & 0 &0 \\
0 & e^{i\alpha_2/2} & 0 \\
0 & 0 &  1\\
  \end{array}\right],
\end{align}
where we have used the following neutrino oscillation data at 
$3\sigma$ level \cite{Forero:2014bxa} given by
\begin{align}
& 0.278 \lesssim s_{12}^2 \lesssim 0.375, \
 0.392 \lesssim s_{23}^2 \lesssim 0.643, \
 0.0177 \lesssim s_{13}^2 \lesssim 0.0294,  
\nn \\
& 
 0.048\ {\rm eV} \lesssim m_{\nu_3} \lesssim  0.051\ {\rm eV},  \
 0.0084\ {\rm eV} \lesssim m_{\nu_2} \lesssim 0.0090\ {\rm eV}, 
   \label{eq:neut-exp}
  \end{align}
and the Dirac phase $\delta$ and Majorana phases $\alpha_{1,2}$ 
are taken to be 
$\delta,\ \alpha_{1,2}\in[0,2\pi]$ in the numerical analysis.
Notice here that one of three neutrino masses is zero because
$f$ is an anti-symmetric matrix, which is symbolized as
  \begin{align}
f&\equiv \left[\begin{array}{ccc}
0 & f_{e\mu} & f_{e\tau} \\
-f_{e\mu}  & 0 & f_{\mu \tau}  \\
-f_{e\tau}  & -f_{\mu\tau}  &  0\\
  \end{array}\right]. \label{Eq:F}
  \end{align}
Therefore, one can rewrite any two components of $f$ in terms of 
experimental values and the remaining component of 
$f$~\cite{Herrero-Garcia:2014hfa}.
Here we select as follows:
\begin{align}
 f_{e\tau}&=\left(\frac{s_{12}c_{23}}{c_{12}c_{13}} + \frac{s_{13} s_{23}}{c_{13}}e^{-i\delta} \right) f_{\mu\tau},\quad 
f_{e\mu}=\left(\frac{s_{12}c_{23}}{c_{12}c_{13}} - \frac{s_{13} s_{23}}{c_{13}}e^{-i\delta} \right)  f_{\mu\tau}.
\label{eq:ys-NH}
  \end{align}
Thus only $f_{\mu\tau}$, {\it which does not contribute to the neutrino mass structure because it is an overall parameter}, 
is an input parameter in our numerical analysis, and we will search for the allowed
region in the parameter space by comparing with the experimental values in
Eqs.~(\ref{eq:exp-Neutmass-NH}).
We assume $g_{i \alpha}$ to be the real matrix for simplicity.

{
{\it The inverted ordering case}; $ D_\nu\equiv (m_{\nu_1},m_{\nu_2},0)$, is also
written as:
\begin{align}
|{\cal M}_{\nu}|
&=|(V_{\rm MNS} D_\nu V_{\rm MNS}^T)|
\nn\\
&\approx \left[\begin{array}{ccc} 
1.00-5.00 & 0.00237-3.83 &0.00256-3.94 \\
* & 0.00279-3.08 &  0.365-2.60 \\
* & * &   0.00500-3.30\\
  \end{array}\right]\times 10^{-11}\ {\rm GeV},\label{eq:exp-Neutmass-IH}
\end{align}
where we have used the following neutrino oscillation data at 
$3\sigma$ level \cite{Forero:2014bxa} given by
\begin{align}
& 0.278 \lesssim s_{12}^2 \lesssim 0.375, \
 0.403 \lesssim s_{23}^2 \lesssim 0.640, \
 0.0183 \lesssim s_{13}^2 \lesssim 0.0297,  
\nn \\
& 
 0.0469\ {\rm eV} \lesssim m_{\nu_1} \lesssim  0.0504\ {\rm eV},  \
 0.0477\ {\rm eV} \lesssim m_{\nu_2} \lesssim 0.0512\ {\rm eV}, 
   \label{eq:neut-exp-IH}
  \end{align}
and $f$ can be rewritten by
\begin{align}
 f_{e\tau}&=-\left(\frac{c_{13} s_{23}}{s_{13}}e^{-i\delta} \right) f_{\mu\tau},
 \quad 
f_{e\mu}=\left(\frac{c_{13} c_{23}}{s_{13}}e^{-i\delta} \right)  f_{\mu\tau}.
\label{eq:ys-IH}
  \end{align}
}

\subsection{Lepton Flavor Violations and Muon Anomalous Magnetic Moment}
{\it $\ell_\alpha \to \ell_\beta \gamma$ process}: First of all, let us consider
the processes $\ell_\alpha \to \ell_\beta \gamma$ at one-loop
level~\footnote{The experimental bounds are summarized in
  Table~\ref{tab:Cif}.}.  The formula for the branching ratio can
generally be written as
\begin{align}
{\rm BR}(\ell_\alpha \to \ell_\beta \gamma)
=
\frac{48\pi^3 C_\alpha \alpha_{\rm em}}{{\rm G_F^2} m_\alpha^2 }\,
  (|(a_R)_{\alpha \beta}|^2+|(a_L)_{\alpha \beta}|^2),
\end{align}
where $\alpha_{\rm em}\approx1/137$ is the fine-structure constant,
$C_{\alpha}\approx(1,1/5)$ for ($\alpha=\mu,\tau$), 
${\rm G_F}\approx1.17\times 10^{-5}$ GeV$^{-2}$ is the Fermi constant, 
and $a_{L/R}$ is respectively given as
\begin{align}
(a_{R})_{\alpha \beta}&\approx
\frac1{(4\pi)^2}\sum_{a=e,\mu,\tau}\sum_{i=1}^3
\left(\frac{ f^\dag_{\beta a} f_{\alpha a} }{12 m^2_{S_1}} m_{\ell_\alpha} +
\frac{ g^\dag_{\beta i} g_{i \alpha}  }{m^2_{S_2}} m_{\ell_\beta} F_I\left[\frac{M_{N_i}^2}{m_{S_2}^2}\right]  \right),
\\
(a_{L})_{\alpha \beta}&=
\frac1{(4\pi)^2}\sum_{a=e,\mu,\tau} \sum_{i=1}^3
\left(\frac{ f^\dag_{\beta a} f_{a \alpha}  }{12 m^2_{S_1}} m_{\ell_\beta} +
\frac{g^\dag_{\beta i} g_{i \alpha}  }{m^2_{S_2}} m_{\ell_\alpha}  F_I\left[\frac{M_{N_i}^2}{m_{S_2}^2}\right]   \right),
 \end{align} 
where 
\begin{align}
&F_I(x)=
\frac{1-6x+3 x^2+2 x^3-6x^2\ln[x]}{6(1-x)^4}.
\end{align}
Once we assume that $m_{\ell_\alpha} \gg m_{\ell_\beta}$, the formula can be simplified to
\begin{align}
{\rm BR}(\ell_\alpha \to \ell_\beta \gamma)\approx
\frac{48\pi^3 C_\alpha \alpha_{\rm em}}{3{\rm G_F^2}(4\pi)^4 }
\left[
\frac{|\sum_{a=e,\mu,\tau} f^\dag_{\beta a} f_{a \alpha}|}{ m^4_{S_1}} +\frac{36} {m^4_{S_2}}
\left|\sum_{i=1}^3 
g^\dag_{\beta i} g_{i \alpha}   F_I\left[\frac{M_{N_i}^2}{m_{S_2}^2}\right]
\right|^2  
\right].
\end{align}

{\it $\mu\mathchar`-e$ conversion}:
The $\mu\mathchar`-e$ conversion rate $R$ can also be written in a similar form as
${\rm BR}(\ell_\alpha \to \ell_\beta \gamma)$ ~\cite{Alonso:2012ji} 
\footnote{In general, those terms proportional to vector-like current:
  $\bar \mu \gamma^\mu (b_L P_L+b_R P_R)e$ via $\gamma/Z$ mediation
  contribute to the $\mu\mathchar`-e$ conversion process. However, these terms are
  negligible in the limit of $M_{N_i},m_{S_2} \gg m_Z,m_{\ell_\alpha}$.
 }
 as
\begin{align}
R&=\frac{\Gamma(\mu\to e)}{\Gamma_{\rm capt}},\quad 
\Gamma(\mu\to e)\approx
C_{\mu e} Z
\left[
\left |(a_R)_{\mu e} \right|^2 + \left|  (a_L)_{\mu e} \right|^2
\right],
\end{align}
where we neglect the contribution from the Higgs-mediated digram due to
the Yukawa coupling suppression, $C_{\mu e}\equiv4\alpha_{\rm em}^5
{Z^4_{\rm eff}|F(q)|^2 m^5_\mu}{}$ and we assume that  $m_{\ell_\alpha},
m_Z \ll  m_{S_2},M_{N_i}$.
The values for $\Gamma_{\rm capt}$, $Z$, $Z_{\rm eff}$, and $F(q)$ 
depend on the type of nuclei, as being shown in Table~\ref{tab:mue-conv}.
One remark from this table is that the sensitivity of Titanium 
will be improved by several orders of magnitude in near future.
Therefore the model testability will increase drastically.

{\it Lepton Universality}: 
A number of lepton-universality experiments (e.g.,
$W$ boson couplings, Kaon decays, pion decays, etc) restrict the coupling of 
$f_{\alpha\beta}$, and the bounds are summarized in 
Table~\ref{tab:lu}~\cite{Herrero-Garcia:2014hfa}.

{\it $\ell_\alpha \to \ell_\beta \ell_\gamma \ell_\sigma$ processes}:  We have
 three-body decay LFV processes at one-loop level with the 
box-type diagram arising from $f$ and $g$,
however these contributions are usually negligibly tiny compared to the 
processes $\ell_\alpha \to \ell_\beta \gamma$. Thus, we do not consider them here, 
but see for details in, e.g, Ref.~\cite{Nishiwaki:2015iqa}
\footnote{In this paper the notation of $f$ should be replaced by $f/2$.}.

\begin{table}[t]
\begin{tabular}{c|c|c|c} \hline
Process & $(\alpha,\beta)$ & Experimental bounds ($90\%$ CL) & References \\ \hline
$\mu^{-} \to e^{-} \gamma$ & $(\mu,e)$ &
	${BR}(\mu \to e\gamma) < 4.2 \times 10^{-13}$ & \cite{TheMEG:2016wtm} \\
$\tau^{-} \to e^{-} \gamma$ & $(\tau,e)$ &
	${BR}(\tau \to e\gamma) < 3.3 \times 10^{-8}$ & \cite{Adam:2013mnn} \\
$\tau^{-} \to \mu^{-} \gamma$ & $(\tau,\mu)$ &
	${BR}(\tau \to \mu\gamma) < 4.4 \times 10^{-8}$ & \cite{Adam:2013mnn}   \\ \hline
\end{tabular}
\caption{Summary for the experimental bounds of the LFV processes 
$\ell_\alpha \to \ell_\beta \gamma$.}
\label{tab:Cif}
\end{table}

\begin{table}[t]
\begin{tabular}{c|c|c|c|c} \hline
Nucleus $^A_Z N$ & $Z_{\rm eff}$ & $|F(-m^2_\mu)|$ & $\Gamma_{\rm capt}(10^6~{\rm sec}^{-1})$ & Experimental bounds (Future bound) \\ \hline
$^{27}_{13} Al$ & $11.5$ & $0.64$ & $0.7054$  & ($R_{Al}\lesssim10^{-16}$~\cite{Hungerford:2009zz})\\
$^{48}_{22} Ti$ & $17.6$ & $0.54$ &$2.59$  & 
$R_{Ti}\lesssim 4.3\times 10^{-12}$~\cite{Dohmen:1993mp}\ 
($\lesssim 10^{-18}$ 
\cite{Barlow:2011zza}) \\
$^{197}_{79} Au$ & $33.5$ & $0.16$ & $13.07$ & $R_{Au}\lesssim7\times 10^{-13}$ ~\cite{Bertl:2006up}  \\ 
$^{208}_{82} Pb$ & $34$ & $0.15$ & $13.45$   & $R_{Pb}\lesssim4.6\times 10^{-11}$~\cite{Honecker:1996zf}  \\ \hline
\end{tabular}
\caption{
Summary for the the $\mu\mathchar`-e$ conversion in various nuclei: 
$Z$, $Z_{\rm eff}$, $F(q)$, $\Gamma_{\rm capt}$, and the bounds on
the capture rate $R$.}
\label{tab:mue-conv}
\end{table}

\begin{table}[t]
\begin{tabular}{c|c|c} \hline
Process  & Experiments & Bound ($90\%$ CL)  \\ \hline
{\rm Lepton/hadron\ universality}  &
$\sum_{q=b,s,d}|V^{\rm exp}_{uq}|^2=0.9999\pm0.0006$:  & $|f_{e\mu}|^2<0.007\left(\frac{m_{S_1}}{{\rm TeV}}\right)^2$   \\ 
${\rm \mu/e\ universality}$ & 
	$\frac{G_\mu^{\rm exp}}{G_e^{\rm exp}}=1.0010\pm0.0009$ & $||f_{\mu\tau}|^2-|f_{e\tau}|^2|<0.024\left(\frac{m_{S_1}}{{\rm TeV}}\right)^2$  \\ 
${\rm \tau/\mu\ universality}$ & 
	$\frac{G_\tau^{\rm exp}}{G_\mu^{\rm exp}}=0.9998\pm0.0013$ & $||f_{e\tau}|^2-|f_{e\mu}|^2|<0.035\left(\frac{m_{S_1}}{{\rm TeV}}\right)^2$  \\ 
${\rm \tau/e\ universality}$ & 
	$\frac{G_\tau^{\rm exp}}{G_e^{\rm exp}}=1.0034\pm0.0015$ & $||f_{\mu\tau}|^2-|f_{e\mu}|^2|<0.04\left(\frac{m_{S_1}}{{\rm TeV}}\right)^2$  \\ \hline
\end{tabular}
\caption{Summary of the lepton universality and the corresponding bounds
on $f_{\alpha\beta}$.}
\label{tab:lu}
\end{table}

{\it Muon anomalous magnetic moment}:
The formula for the muon $g-2$ can be written in terms of $a_L$ and $a_R$, 
and simplified as follows:
\begin{align}
\Delta a_\mu\approx -{m_\mu}(a_R+a_L)_{\mu \mu}
\approx
-\frac{m^2_\mu}{96\pi^2}\sum_{a=e,\mu,\tau} \sum_{i=1}^3
\left(\frac{ f^\dag_{\mu a} f_{a \mu}  }{m^2_{S_1}} +6
\frac{g^\dag_{\mu i} g_{i \mu}  }{m^2_{S_2}}  F_I\left[\frac{M_{N_i}^2}{m_{S_2}^2}\right]   \right)
.\label{damu}
\end{align}
Notice here that this contribution to the muon $g-2$ is negative, yet it
is negligible compared to the deviation in the experimental 
value~${\cal O}(10^{-9})$~\cite{Bennett:2006fi}.

\subsection{Dark Matter}
{\it Relic density}:
Here we identify $N_3$ as the DM candidate and denote its mass by 
$M_{N_3}\equiv M_X$.
Also, we include the coannihilation system with [$N_{1},N_{2}, S_2^\pm$]
in order to suppress the relic density to satisfy the experimental value.
We adopt the approximation in relative-velocity expansion up to the $p$-wave.
The relic density is then given by
\begin{align}
\Omega h^2\approx \frac{1.07\times10^9 }
{\sqrt{g^*} M_P \int_{x_f}^\infty dx
\left[\frac{a_{\rm eff}}{x^2}+\frac6{x^3} (b_{\rm eff}-\frac{a_{\rm eff}}{4})\right] },
\end{align}
where $g^*\approx100$, $M_P\approx 1.22\times 10^{19}$, $x_f\approx25$, and
each of the coefficients for s-wave and p-wave can be written 
in terms of summations over several modes as follows:
\begin{align} 
\frac{g_{\rm eff}^2}{4} a_{\rm eff}& \simeq a (N_i N_j\to\ell\bar\ell)+  a(N_i S_2^+\to \ell^+\gamma)+  a(N_i S_2^+\to\ell^+ Z)\nn\\
&+ a( S_2^+ S_2^- \to2\gamma)+  a( S_2^+ S_2^- \to 2Z)  +  a( S_2^+ S_2^- \to2h)+
a( S_2^+ S_2^- \to t\bar t), \\
\frac{g_{\rm eff}^2}{4} b_{\rm eff}& \simeq  b( X\bar X \to\ell\bar\ell)
 +b(N_i N_j\to\ell\bar\ell)+  b(N_i S_2^+\to \ell^+\gamma)+  b(N_i S_2^+\to\ell^+ Z)\nn\\
& +  b( S_2^+ S_2^- \to2\gamma)+  b_{}( S_2^+ S_2^- \to 2Z)  +  b( S_2^+ S_2^- \to2h)+
b( S_2^+ S_2^- \to t\bar t).
\end{align}
Furthermore, $a(b)_{ij\to k\ell}$ is given in terms of the cross 
section expanded by the relative velocity $v_{\rm rel}$ as follows:
\begin{align}
(\sigma v_{\rm rel})(ij\to k\ell)&\approx 
\frac{\sum_{i,j}}{32 \pi^2 s_{ij}}\sqrt{1-\frac{(m_k+m_\ell)^2}{s_{ij}}} \int d\Omega |\bar M(ij\to k\ell)|^2
(1+\Delta_i)^{2/3} (1+\Delta_j)^{2/3} e^{-x(\Delta_i+\Delta_j)}
\nn\\
&\approx\sum_{i,j}[a(ij\to k\ell)+b(ij\to k\ell) v_{\rm rel}^2](1+\Delta_i)^{2/3} (1+\Delta_j)^{2/3} e^{-x(\Delta_i+\Delta_j)},
\end{align}
where $\Delta_i\equiv \frac{m_i-M_X}{M_X}$, $d\Omega=2\pi\int_0^{\pi}d\theta \sin\theta$, and
\begin{align}
&g_{\rm eff}\equiv \sum_i g_i (1+\Delta_i)^{3/2}e^{-x\Delta_i}\nn\\
&=2 \left[(1+\Delta_X)^{3/2}e^{-x\Delta_X}+ (1+\Delta_{N_2})^{3/2}e^{-x\Delta_{N_2}}
+ (1+\Delta_{N_1})^{3/2}e^{-x\Delta_{N_1}} + (1+\Delta_{S_2})^{3/2}e^{-x\Delta_{{S_2}}}\right].
\end{align}
Now  the explicit forms for $a$ and $b$ should be written down, 
where the $a(N_i\bar N_j\to \ell\bar\ell)$, 
 $a(N_i\bar N_j\to \ell\bar\ell)$, and $b(X\bar X\to \ell\bar\ell)$ can be
found in \cite{Ahriche:2013zwa}.
Thus, we write down the other modes 
$N_i S_2^+\to f_1f_2^*$ and $S_2^+ S_2^-\to f_1f_2^*$ as mass invariant squared:
\begin{align}
&|\bar M(N_i S_2^+ \to \ell_\alpha \gamma)|^2\approx \sum_{i=1}^3 \sum_{\alpha=e,\mu,\tau} \left|\frac{e g^\dag_{i \alpha}}{s-m^2_{\ell_\alpha}}\right|^2
\left(2(p_1\cdot k_1+p_2\cdot k_1)(M^2_i+p_1\cdot p_2)-s (p_1\cdot k_1) \right),\\
&|\bar M(N_i S_2^+ \to \ell_\alpha Z)|^2\approx  
\sum_{i=1}^3 \sum_{\alpha=e,\mu,\tau}^3\left|\frac{s_{{\theta_w}}^2 g_2 g^\dag_{i \alpha}}{c_{{\theta_w}}(s-m^2_{\ell_\alpha})}\right|^2
\left[
(m^2_{S_2}-M^2_i)(p_1\cdot k_1)\right.
\nn\\&
\left.
-2\{(m^2_{S_2}-M^2_i)(p_1\cdot k_2)
-2 (M^2_i + p_1\cdot p_2)(p_2\cdot k_2) \} \frac{(k_1\cdot k_2)}{m_Z^2}
+ 
(M^2_i + p_1\cdot p_2)(p_2\cdot k_1)
 \right],\\
& |\bar M(S_2^+S_2^- \to 2\gamma)|^2\approx 
\frac{(4\pi\alpha_{\rm em})^2}2 G^{(\gamma)}_{\mu\nu} G^{(\gamma)\mu\nu},\\
& |\bar M(S_2^+S_2^- \to 2Z)|^2\approx
\frac12\left(-g_{\mu\alpha}+\frac{k_{1\mu} k_{1\alpha}}{m_Z^2}\right)
\left(-g_{\nu\beta}+\frac{k_{1\nu} k_{1\beta}}{m_Z^2}\right)
 G^{(Z)\mu\nu} G^{(Z)\alpha\beta},\\
&|\bar M( S_2^+S_2^- \to 2h)|^2\approx 
\left|\lambda_{\Phi S_2}+\frac{3v^2 \lambda_{\Phi}\lambda_{\Phi S_2} }{4(s-m_h^2)}+\frac{(\lambda_{\Phi S_2}v)^2}{4}
\left(\frac{1}{t-m^2_{S_2}} + \frac{1}{u-m^2_{S_2}}\right)
 \right|^2,\\
&|\bar M( S_2^+S_2^- \to t\bar t)|^2\approx \nn\\
&{\rm Tr}\left[
(\slashed k_1+m_t)[A+B (\slashed p_1+ \slashed p_2)+C (\slashed p_1+ \slashed p_2)\gamma_5](\slashed k_2-m_t)]
[A+B (\slashed p_1+ \slashed p_2)-C \gamma_5(\slashed p_1+ \slashed p_2)]
\right],
\end{align}
where {$s_{\theta_w} (c_{\theta_w}) \equiv \sin \theta_w (\cos \theta_w)$ denotes the Weinberg angle with $\sin^2 \theta_w = 0.23$,} 
\begin{align}
G^{(\gamma)}_{\mu\nu}&\equiv
g_{\mu\nu}+\frac{(2p_1-k_1)_\mu (p_2-p_1+k_1)_\nu}{t-m^2_{S_2}} +\frac{(2p_1-k_2)_\nu (p_2-p_1+k_2)_\mu}{u-m^2_{S_2}} ,\\
G^{(Z)}_{\mu\nu}&\equiv
g_{\mu\nu}\left[\frac{g_2^2 s_{{\theta_w}}^4}{c_{{\theta_w}}^2}+\frac{\lambda_{\Phi S_2} m^2_Z}{s-m^2_h} \right]
+\left[\frac{g_2 s_{{\theta_w}}^2}{c_{{\theta_w}}}\right]^2
\left[\frac{(2p_1-k_1)_\mu (p_2-p_1+k_1)_\nu}{t-m^2_{S_2}} +\frac{(2p_1-k_2)_\nu (p_2-p_1+k_2)_\mu}{u-m^2_{S_2}}\right] ,\\
A &\equiv \frac{\lambda_{\Phi S_2} m_t v}{s-m_h^2},\quad
B\equiv \frac{2 e^2}{3 s^2} +\left(\frac14-\frac23 s^2_{{\theta_w}}\right)\frac{s^2_{{\theta_w}}g_2^2}{c^2_{{\theta_w}} m^2_Z},\quad
C\equiv -\frac{s^2_{{\theta_w}} g_2^2}{4 c_{{\theta_w}}^2 m_Z^2},
\end{align}
and  $p_{1/2}$ are the initial momenta and $k_{1/2}$ are the final momenta. 
In appendix, we explicitly show the formulas of Mandelstam variables and 
the scalar products in the $v_{\rm rel}$-expanded form.
Note that the s-wave contributions are suppressed since they are 
proportional to the square of down-type quark mass.
In our numerical analysis below, we use the current 
experimental range approximately as 
$0.11\le \Omega h^2\le 0.13$~\cite{Ade:2013zuv}. 

{\it Direct detection}:
When the masses among $N_i$ are degenerate
\footnote{Typical mass difference is within the scale of the order of 100 keV.},
DM inelastically interacts with nucleon through $\gamma/Z$ at 
one-loop level~\cite{Schmidt:2012yg}. 
However it does not reach the sensitivity of current detectors 
such as LUX~\cite{Akerib:2016vxi}.

\subsection{Collider physics}
The collider signatures for the KNT model were considered in Ref.~\cite{seto}
for linear colliders.  We shall briefly highlight here. 
The lightest RH neutrino $N_3$ is the dark matter candidate, while 
the other RH neutrinos $N_{1,2}$ and the charged boson $S^+_2$ are slightly
heavier because of the requirement of coannihilation.

At $e^+ e^-$ colliders, one can produce $e^+ e^- \to N_3 N_{1,2}$ followed
by the decays of $N_{1,2} \to N_3 \ell^+ \ell'^-$, which gives rise to 
a final state of a pair of charged leptons (not necessarily the same
flavor) plus missing energies.
One can also consider the pair production of $S_2^+ S_2^-$ via
$e^+ e^- \stackrel{\gamma^*,Z^*}{\longrightarrow} S_2^+ S_2^-$. 
Note that the $t$-channel
diagram with an exchange of a RH neutrino is suppressed by the mass of
the RH neutrino. The $S_2^\pm$ so produced will decay into $\ell^\pm N_3$,
and so the final state consists of a pair of charged leptons
(again not necessarily the same flavor) and missing energies.

The decay of $N_{1,2}$ is analogous to the heavier neutralinos 
$\tilde{\chi}_{2,3}^0$ in the minimal supersymmetric standard model (MSSM),
which can then give a pair of charged leptons plus missing energies. 
On the other hand, the decay of $S_2^\pm$ is analogous to the slepton
in MSSM.  Therefore, the limits from $e^+ e^-$ colliders mainly
come from LEP2, and the limits are roughly \cite{pdg}, without
taking any assumption on the underlying particle theory,
\[
 M_{N_{1,2,3}}, m_{S^\pm_2} \alt 85 - 105 \; {\rm GeV} \;.
\]

At hadron colliders, the leading order production process is the Drell-Yan
process $pp \stackrel{\gamma^*,Z^*}{\longrightarrow} S_2^+ S_2^-$,
followed by the decays of the $S_2^\pm \to \ell^\pm N_3$. The final state
consists of a pair of charged leptons (again not necessarily of the same
flavor) plus missing energies. 
Such a signature is possible at the LHC and indeed the final state is 
similar to the direct production of a chargino pair at the LHC, 
in which each chargino can decay into the lightest neutralino and 
a charged lepton.
Thus, the final state consists of a pair of charged leptons whose flavors
can be different, plus missing energies.
For example, the ATLAS Collaboration has searched for the same and different
lepton flavors plus missing energies at the LHC, using the channel
$pp \to \tilde{\chi}^+_1 \tilde{\chi}^-_1 \to (l^+ \nu_l \tilde{\chi}^0_1 )
(l'^- \bar{\nu}_{l'} \tilde{\chi}^0_1 )$  \cite{atlas-emu}.
The best mass limit on the chargino is 
$m_{\tilde{\chi}^\pm_1} \agt 470$ GeV for $m_{\tilde{\chi}^0_1} = 0-100 $ GeV,
but for heavier $m_{\tilde{\chi}^0_1}$ the mass limit for 
$m_{\tilde{\chi}^\pm_1}$ becomes much weaker because of the soft leptons. 
Such mass limits have no relevance to the mass of $S_2^\pm$ that we 
are considering here.

\begin{figure}[h]
\includegraphics[width=8cm,clip]{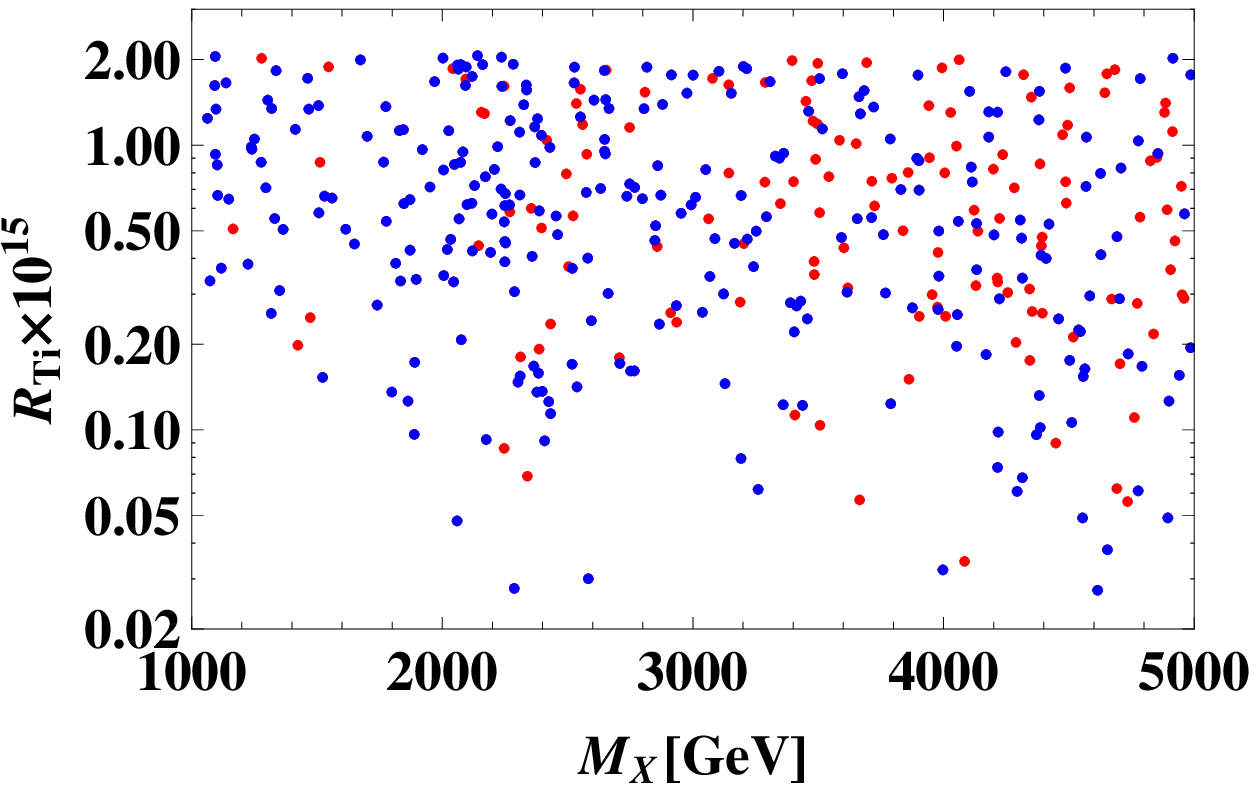}~
\includegraphics[width=8cm,clip]{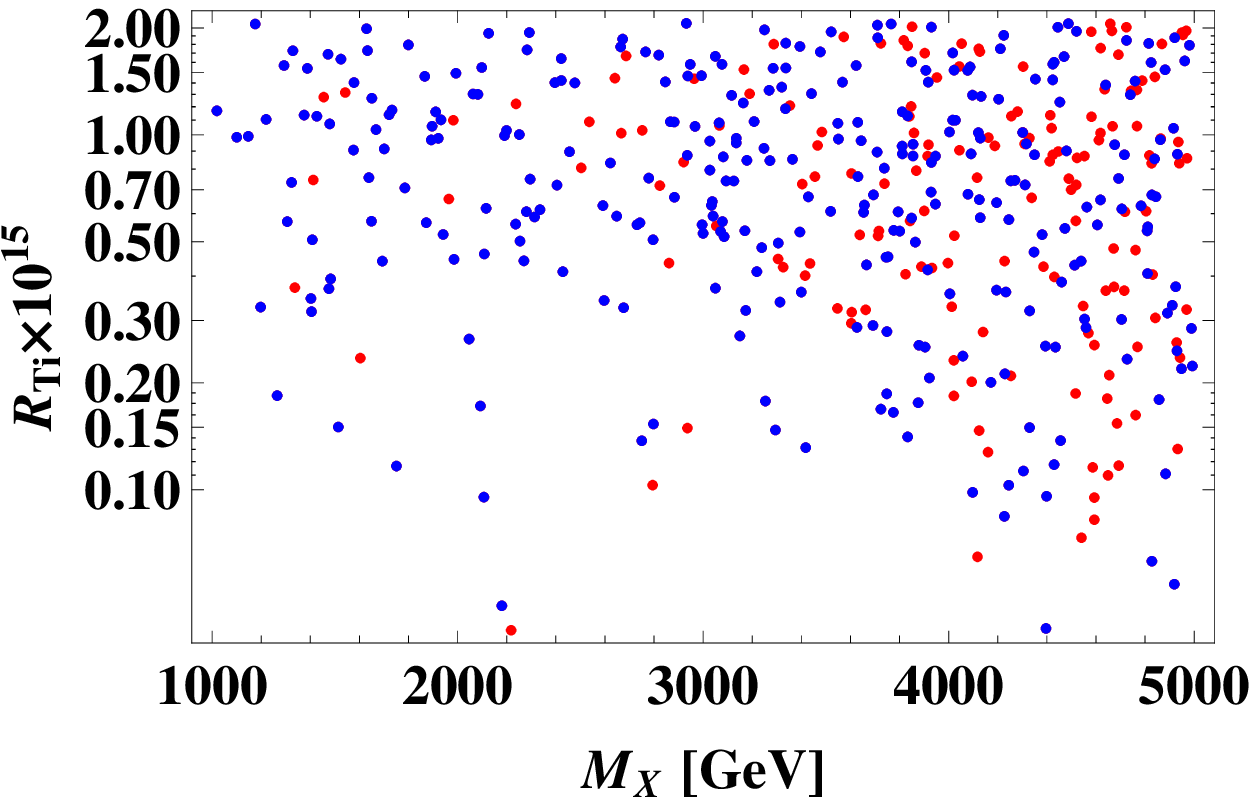}\\\vspace{0mm}
\includegraphics[width=8cm,clip]{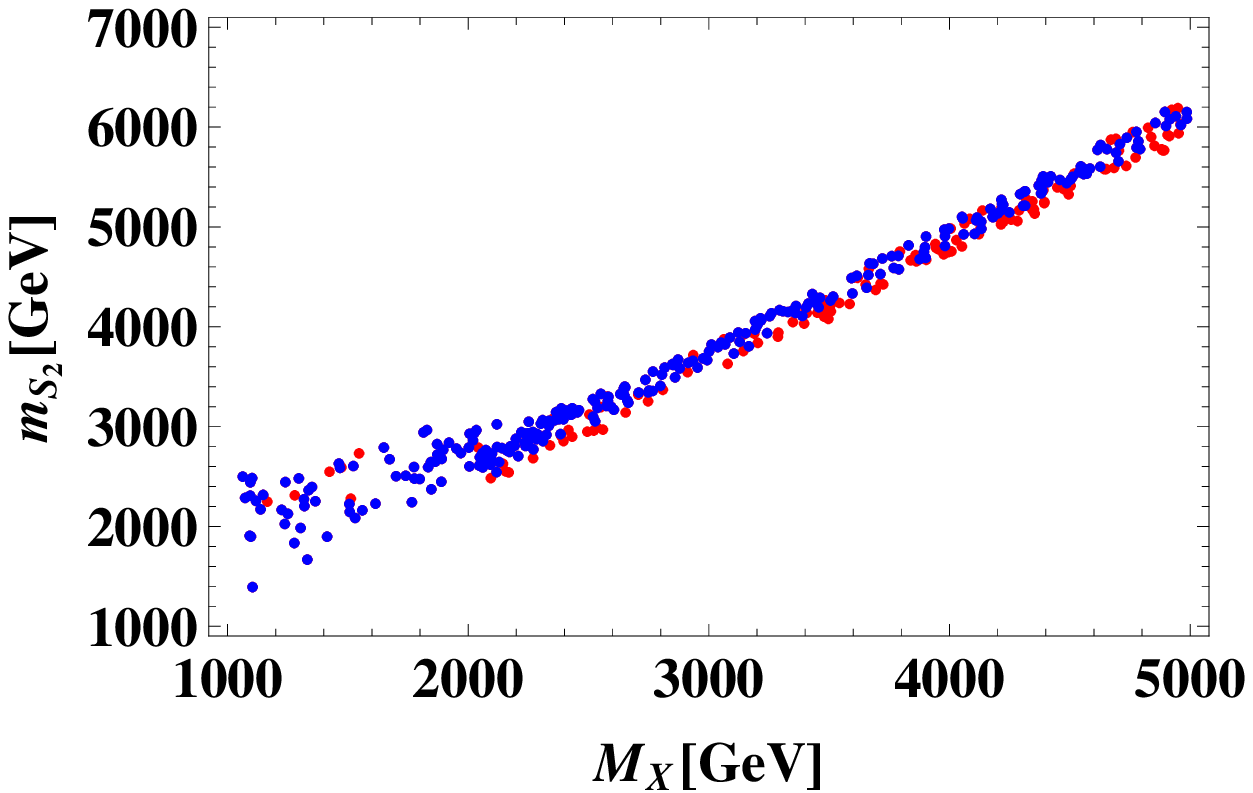}~
\includegraphics[width=8cm,clip]{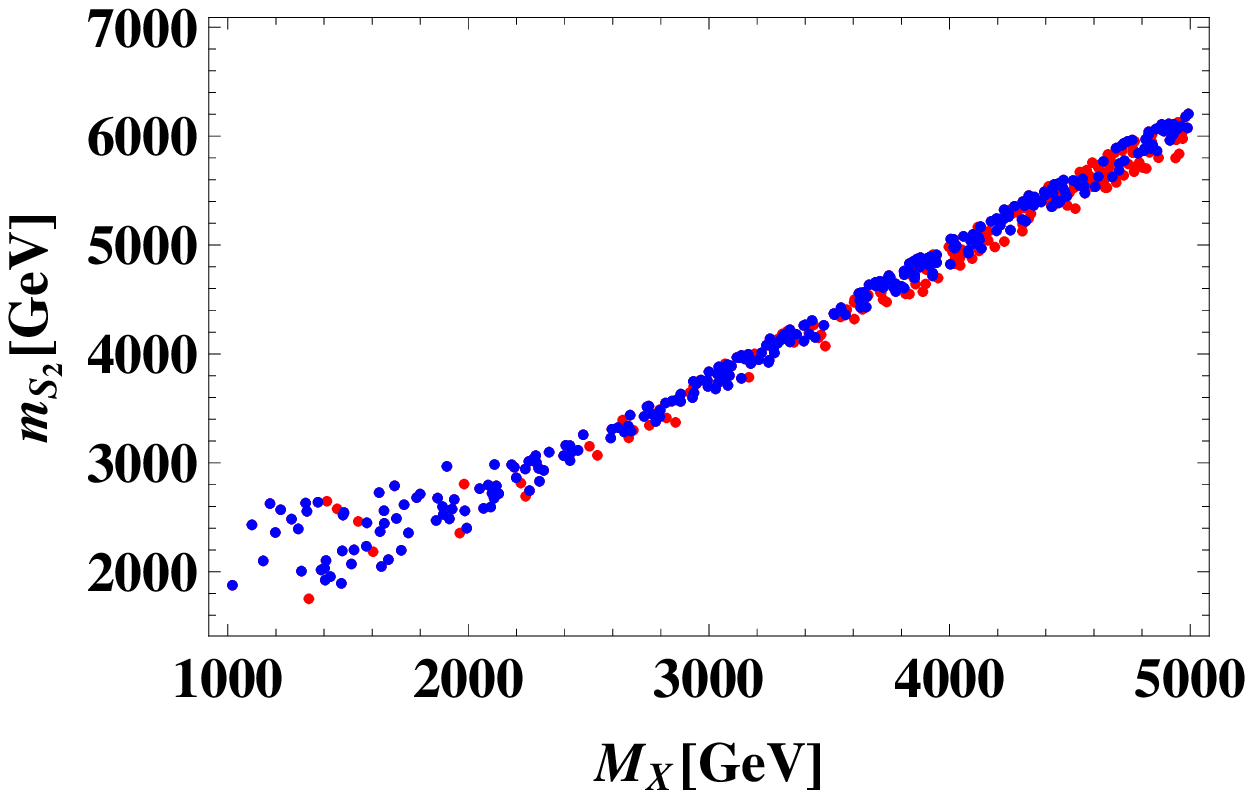}\\\vspace{0mm}
\includegraphics[width=8cm,clip]{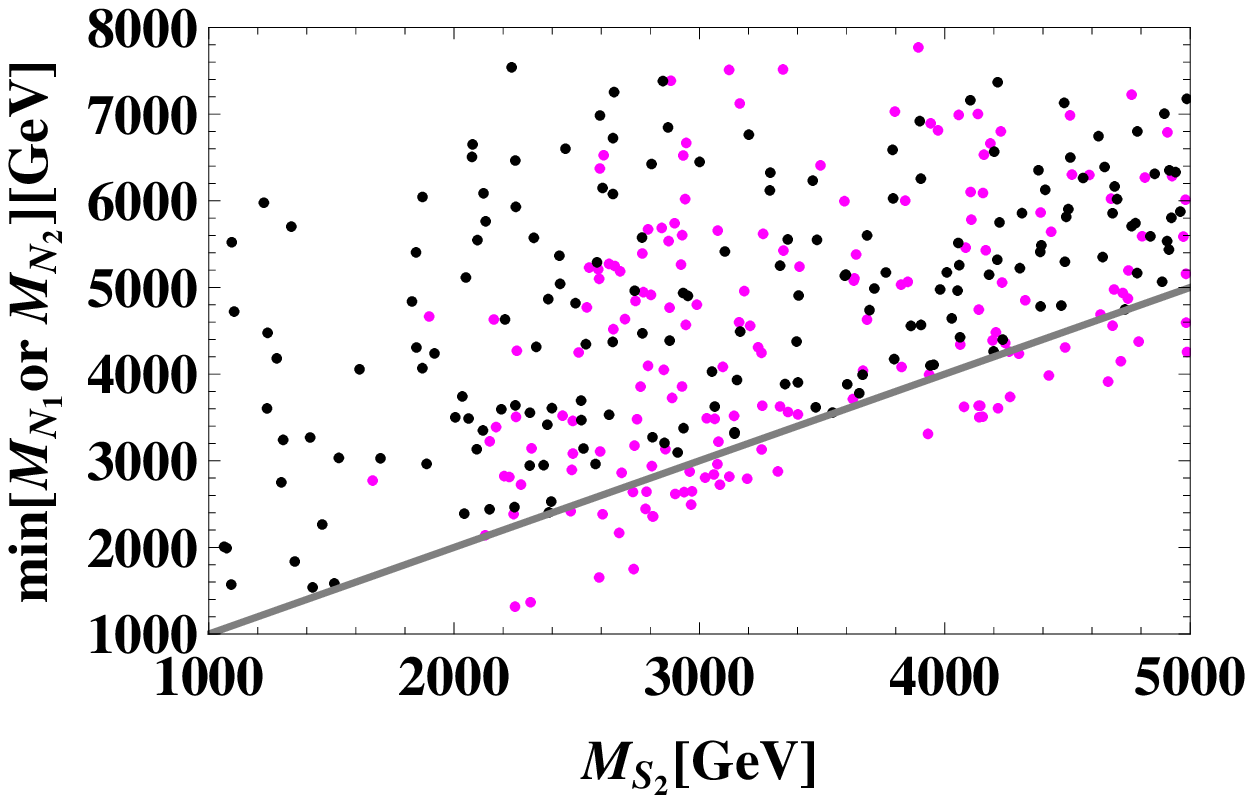}~
\includegraphics[width=8cm,clip]{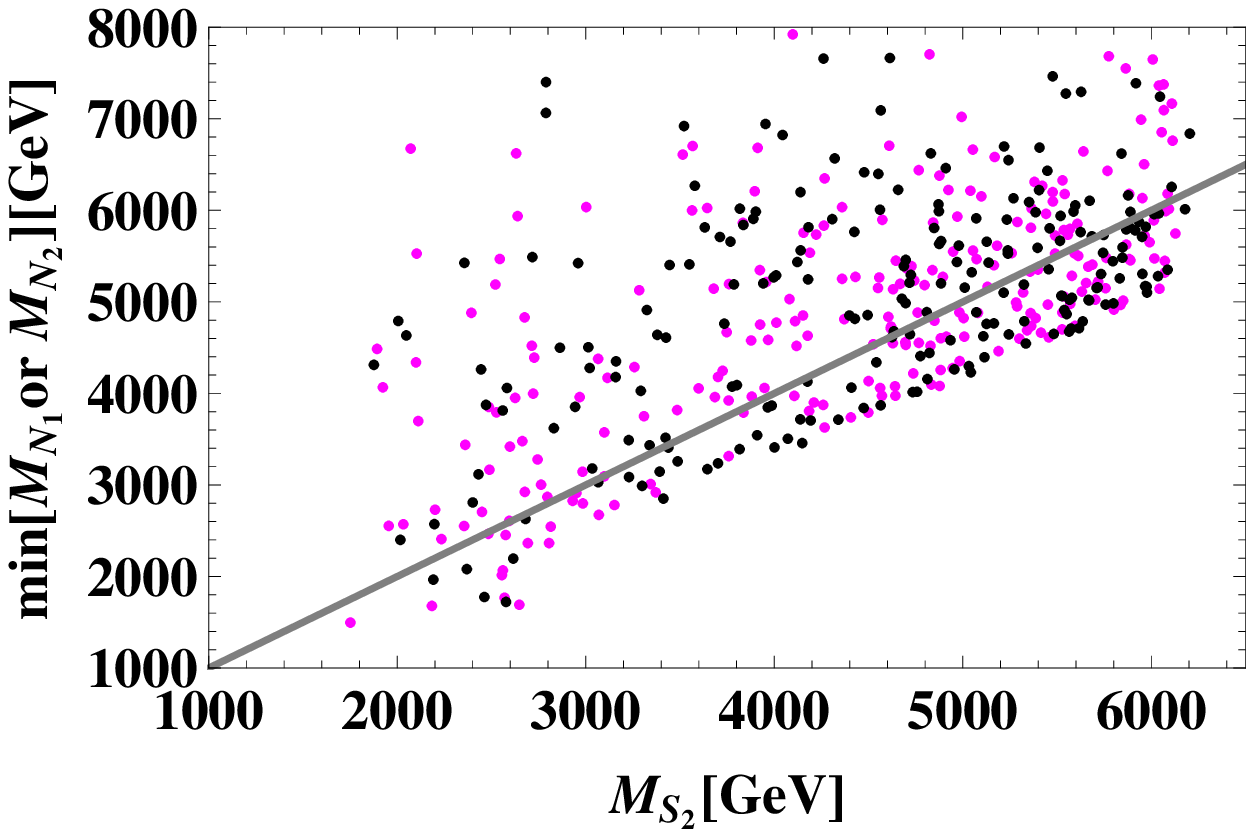}
\caption{
The left panels represent the case of NH while the right ones represent 
the case of IH.
The top panels show the $\mu \mathchar`- e$ conversion capture rate of 
Titanium versus the DM mass.
The middle panels show the mass of $S_2$ versus the DM mass.
The bottom panels show the masses of $N_{1,2}$ versus the mass of $S_2$. 
Here the red points represent the allowed points in the coannihilation 
region, while the blue ones show the annihilation region for the top and 
middle panels.
In the bottom panels,
the gray solid line indicates the equality 
$M_{S_2} = {\rm min} [M_{N_1}~{\rm or}~M_{N_2}]$.
The magenta points represent the range of $M_{N_1} \leq M_{N_2}$, while
the black ones show the range of $M_{N_2} \leq M_{N_1}$.
}\label{fig:nums-1}

\end{figure}

\begin{figure}[h]
\includegraphics[width=8cm,clip]{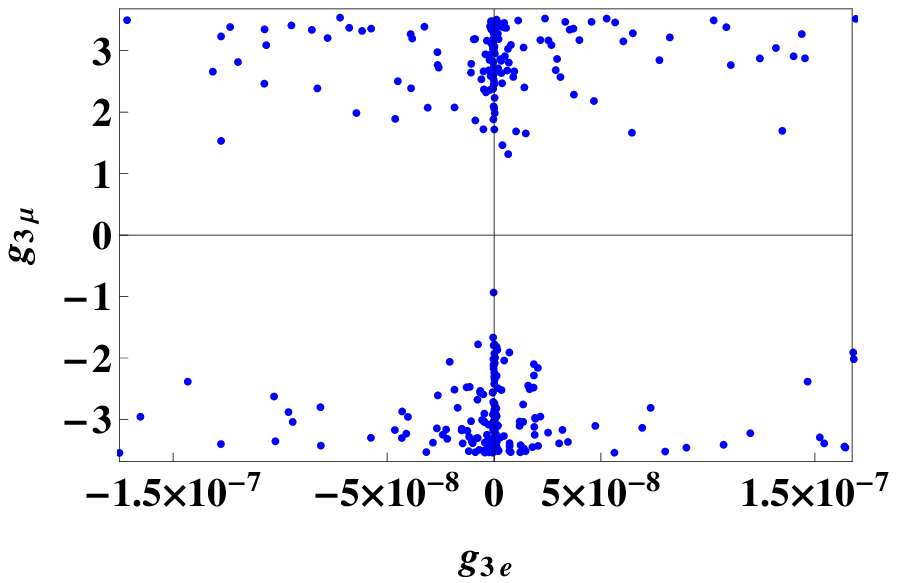}~
\includegraphics[width=8cm,clip]{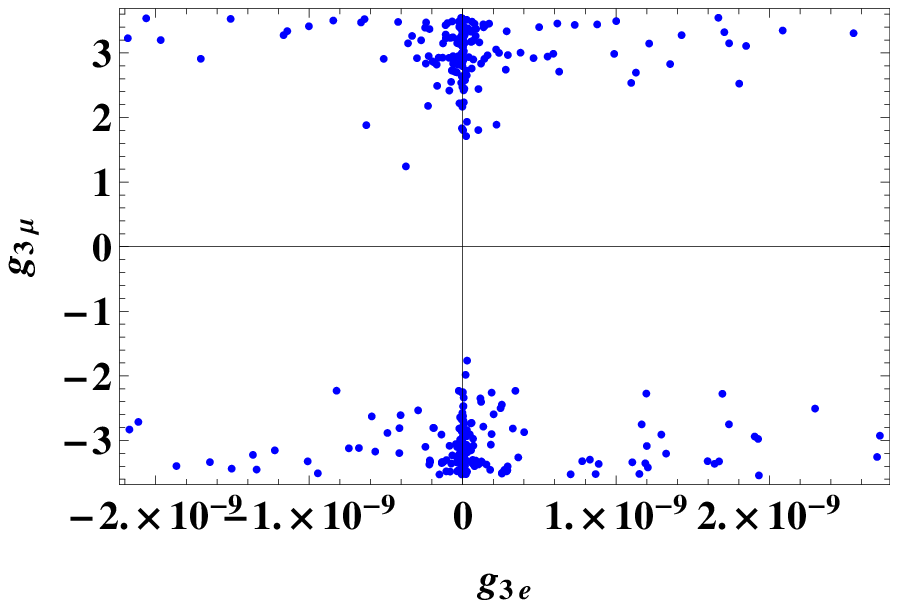}\\\vspace{0mm}
\includegraphics[width=8cm,clip]{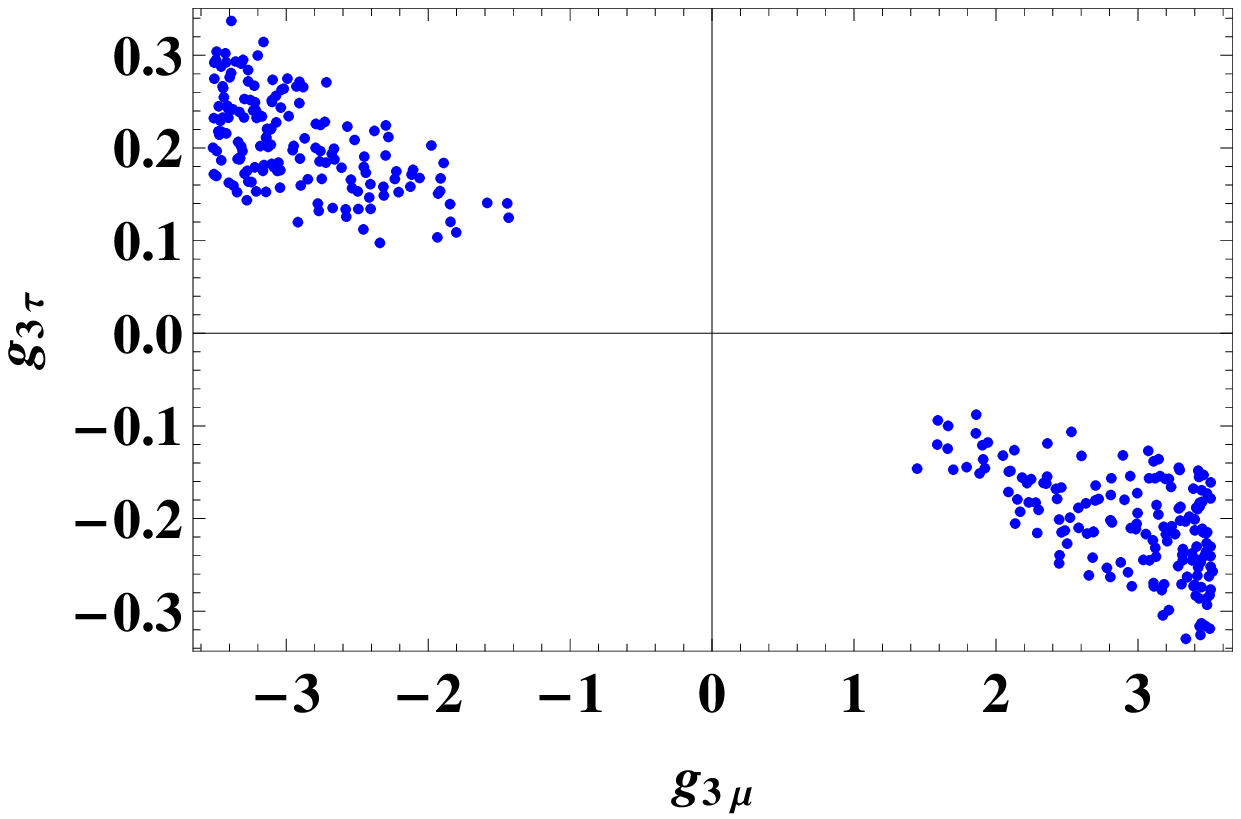}~
\includegraphics[width=8cm,clip]{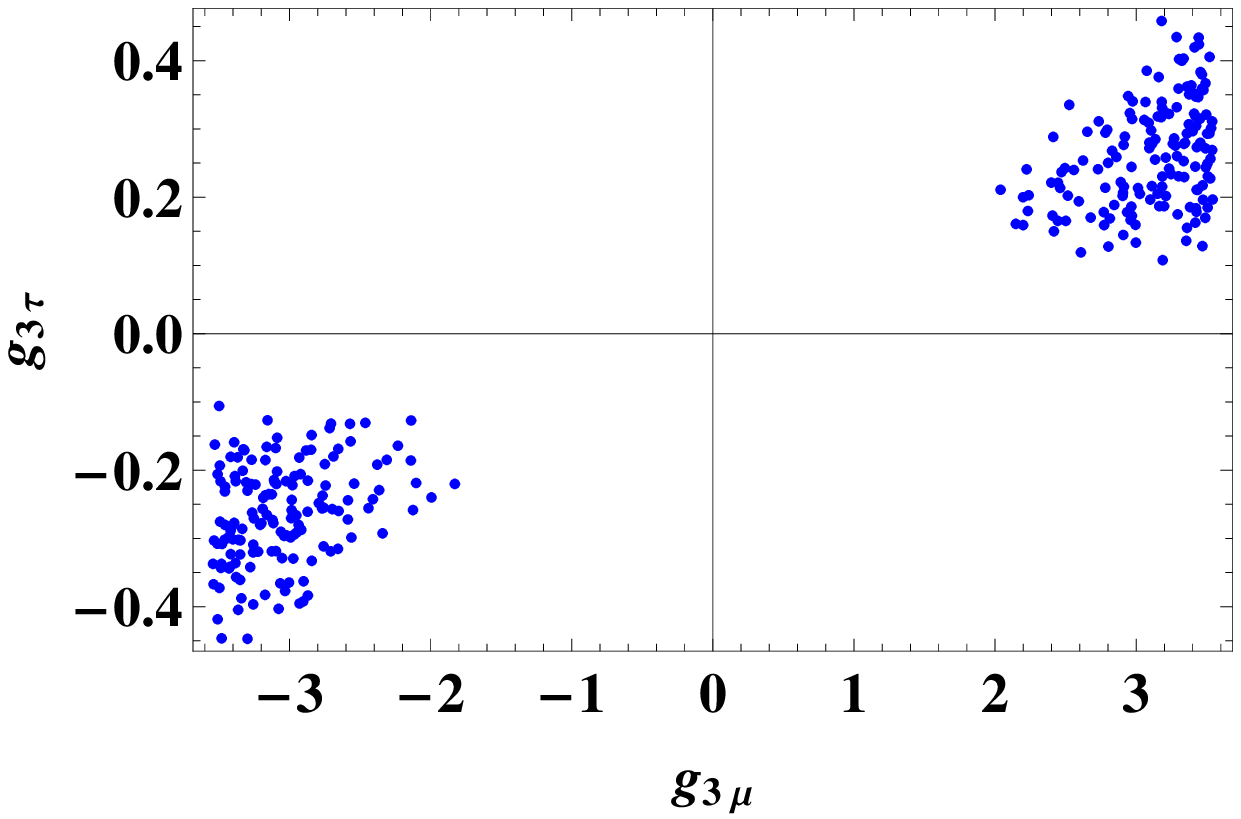}
\caption{
Correlations between $g_{3\mu}$ and $g_{3e}$ (upper panels), and 
between between $g_{3\tau}$ and  $g_{3\mu}$ (lower panels).
The left and right panels correspond to the NH and IH cases, respectively. 
}
\label{fig:nums-2}
\end{figure}

\subsection{Numerical analysis \label{sec:numerical}}

In this subsection, we show the allowed parameter space region that
satisfies all the
constraints. {\it i.e.}, vacuum stability for charged bosons, neutrino
oscillations, LFVs, and the relic density of DM, {for both
normal and inverted cases}. At the first step, we
fix some parameters independent of neutrino mass hierarchy as $\lambda_{0}=4\pi$, $\delta=3\pi/2$, and
$\lambda_{S_{1(2)}}=\pi$, where $\lambda_{0}$ is chosen at the limit
of perturbativity, which is in favor of inducing sizable neutrino masses. 
For other dimensionless couplings we take the following range: 
\begin{align}
({|f_{\mu\tau}}|,|g|) \in [0, \sqrt{4\pi}],\quad
\quad
( \lambda_{\Phi S_1} ,\  \lambda_{\Phi S_2}) \in [0, 0.1], 
\qquad \ \alpha_{1,2} \in [0, 2\pi].
\end{align}

{
Also we take our relevant input mass parameters in the following ranges:
\begin{eqnarray}
1000 \leq &M_X& \leq 5000~{\rm GeV}, \\
M_X \leq &M_{N_i}& \leq 7500~{\rm GeV}, \\
2000 \leq &M_{S_1}& \leq 8000\ {\rm GeV}~~{\rm for}~~M_X \leq 3000~{\rm GeV}, \\
M_X \leq &M_{S_1}& \leq 8000\ {\rm GeV}~~{\rm for}~~M_X \geq 3000~{\rm GeV}, \\
M_X \leq &M_{S_2}& \leq 5000\ {\rm GeV}~~{\rm for}~~M_X \leq 3000~{\rm GeV}, \\
M_X \leq &M_{S_2}& \leq 1.5 M_X\ {\rm GeV}~~{\rm for}~~M_X \geq 3000~{\rm GeV},
\label{Eq:NH}
\end{eqnarray}
for NH and 
\begin{eqnarray}
1000 \leq &M_X& \leq 5000~{\rm GeV},\\
M_X \leq &M_{N_i}& \leq 7500~{\rm GeV},\\
3000 \leq &M_{S_1}& \leq 8000\ {\rm GeV}~~{\rm for}~~M_X \leq 3000~{\rm GeV},\\
M_X \leq &M_{S_1}& \leq 8000\ {\rm GeV}~~{\rm for}~~M_X \geq 3000~{\rm GeV},\\
M_X \leq &M_{S_2}& \leq 5000\ {\rm GeV}~~{\rm for}~~M_X \leq 3000~{\rm GeV},\\
M_X \leq &M_{S_2}& \leq 1.1 \left( M_X + 1000 \right) {\rm GeV}~~{\rm for}~~M_X \geq 3000~{\rm GeV},
\label{Eq:IH}
\end{eqnarray}
for IH, respectively.
Then the numerical results are shown in Figs.~~\ref{fig:nums-1} and  Figs.~~\ref{fig:nums-2}, in which
{
all those on the left side represent}
 the case of NH, while { those on the right side represent}
 the case of IH.

The top {panels of Figs.~\ref{fig:nums-1} represent} the 
$\mu \mathchar`- e$ conversion capture rate of Titanium in terms of the DM mass.
{They suggest that the favorable region for $\mu \mathchar`- e$ 
conversion is relatively larger than that of the future experiment }
${\cal O}(10^{-18})$.
Thus, {in the future experiment, PRISM for instance, one 
can search for the whole region of the relevant parameter space.}
{
Comparing between the NH and IH cases, the NH case tends to 
have a smaller valid parameter space region than the IH case.
The upper limit in both cases arises from the constraint of 
$\mu\to e\gamma$ and are the same because of the same structure as seen 
in Eq.~(21) and Eq.(22). }
Here the red points represent the allowed region in coannihilation, while 
the blue ones are the allowed ones in annihilation {only,}
although both are widely allowed. 
{
The middle panels of Figs.~\ref{fig:nums-1} show the mass of $S_2$ versus
the DM mass.} It implies that 
the mass of $S_2$ is rather degenerated to the mass of DM 
{in order to realize the correct relic} abundance of the DM. 
This is 
{as expected~\cite{Ahriche:2013zwa},
 because $S_2$ is directly involved in the annihilation cross section 
of the relic density.} Thus, considering the coannihilation among 
$S_2$ as well as $N_{1,2}$ is important in the higher DM mass region. 
The red points represent the allowed ones in the coannihilation and the
blue ones are annihilation only.
{
The red ones require somewhat more degenerate mass between $m_{S_2}$ and  $M_X$.}

 The bottom panels of Figs.~\ref{fig:nums-1} show the masses of $N_{1,2}$ 
versus the mass of $S_2$, where
the magenta points represent the range of $M_{N_1} \leq M_{N_2}$ and 
the black ones for the range of $M_{N_2} \leq M_{N_1}$. 
{
They suggest that a wider mass range of $N_{1,2}$ is allowed when 
coannihilation is included.}
In case of NH, only the $N_1$ can be lighter than the mass of $S_2$, 
which is depicted as the allowed points (magenta) below the gray line. 
On the other, the IH allows both hierarchies: $M_{N_1} \leq M_{N_2}$ and
 $M_{N_2} \leq M_{N_1}$. 
It is due to the mass spectrum of active neutrinos. 
This is one of the remarkable differences between NH and IH.

In Figs.~\ref{fig:nums-2}, we plot the correlations among the couplings $g$,
where we specify only the components having the remarkable property, 
which arises from the components $g_{3\ell}$ $(\ell=e,\mu,\tau)$ due to 
being related to the relic density of DM as well as LFVs. 
Therefore, the third row components should be rather large from the 
relic density requirement while LFVs have to be satisfied. 
The other components are widely allowed in whole the ranges that we initially
fix, and therefore only the LFVs have to be satisfied.
The left panels in Fig.~\ref{fig:nums-2} represent the case of NH, 
while the right ones represent the case of IH.
The top panels show the correlation between $g_{3\mu}$ and  $g_{3e}$, and
the bottom ones show the correlation between $g_{3\tau}$ and  $g_{3\mu}$.
Among the $g_{3\ell}$s, each mode of the LFV $\mu\to e\gamma$, 
$\tau\to e\gamma$, and $\tau\to \mu\gamma$ is essentially proportional 
to $g^*_{3e}g_{3\mu}$, $g^*_{3e}g_{3\tau}$, and $g^*_{3\mu}g_{3\tau}$, respectively.
On the other hand, the annihilation cross section that explains 
the relic density requires larger $g_{3\mu}$ and $g_{3\tau}$.
In order to suppress the LFVs of $\mu\to e\gamma$ and $\tau\to e\gamma$ 
below the experimental limits, a tiny $g_{3e}$ is favored to compensate for 
the large components of $g_{3\mu}$ and/or $g_{3\tau}$. 
Obviously, the value of $g^*_{3\mu}g_{3\tau}$ should be constrained in 
order to satisfy the remaining LFV bound on $\tau\to\mu\gamma$.

{
However, one might be (a little bit) skeptical about the size difference
between $g_{3\mu}$, and $g_{3\tau}$ in Figs.~\ref{fig:nums-2}, 
because $g_{3\mu}$ should be more constrained than $g_{3\tau}$ as 
$\mu\to e\gamma$ is more stringent. 
In order to answer this question, one has to scrutinize the structure of 
the active neutrino masses, which is given by Eq.~(\ref{Eq:act_mass}) 
with the structure of Eq.~(\ref{Eq:F}). 
Since the diagonal elements of Eq.~(\ref{Eq:F}) are zero, 
the typical magnitude of the active neutrino mass matrix elements is 
given by the combination of $m_{\ell}g^\dag g^* m_\ell$. 
Furthermore, the typical order of the right-lower elements in the two-by-two 
matrix needs to be the same in order to realize the almost maximal 
mixing of $\theta_{23}$. 
As a result, $g_{3 \mu}$ is required to be one order larger than $g_{3 \tau}$ 
to compensate the mass difference between muon and tau lepton. 
The quantitative results of this point appear 
in Eqs.~(\ref{eq:exp-Neutmass-NH}) and (\ref{eq:exp-Neutmass-IH}) 
for both hierarchy cases. 
}
}

\section{Conclusions}
{
Motivated by a recent result of T2K on the fixed $CP$-odd phase $\delta=3\pi/2$
\cite{t2k-talk}, we have investigated the possibility of accommodating the
$CP$-odd phase $\delta$ in the framework of the Krauss-Nasri-Trodden (KNT)
model supplemented by a total of 3 right-handed neutrinos of mass TeV.
We have analyzed the neutrino oscillation data, lepton-flavor violations, 
and the DM relic density in a coannihilation system including 
additional charged scalars $S_2^\pm$ and heavier right-handed neutrinos 
$N_{2,1}$ in the setup, 
and found the allowed parameter regions that satisfy all the constraints above.

Here we would like offer a few interesting observations as follows.
\begin{enumerate}   
\item
The typical {$\mu$-$e$} conversion rate is at 
the order of $10^{-16}\sim10^{-15}$, which is below the current bound 
by about {four orders of} magnitude. 
{Such conversion rates can be tested at the future 
experiment of $R_{Ti}$} as shown in the top of Figs.~2.
Also, the minimal values are at the order of $10^{-18}$, which imply 
that it could completely be tested by the future experiment of 
COMET Collaboration~\cite{Hungerford:2009zz} and PRISM~\cite{Barlow:2011zza}.
\item 
The mass of $S_2$ lies very close to the mass of DM in order to realize 
the correct abundance of the DM as shown in the middle panels 
of Figs.~\ref{fig:nums-1}, because $S_2$ is directly related to the 
annihilation cross section of the relic density. This result was also 
favored by Ref.~\cite{Ahriche:2013zwa}.
Thus, the coannihilation system among $S_2$ as well as $N_{1,2}$ becomes 
important in the higher DM mass region. 
 \item One could locate the difference between NH and IH, by looking 
for the degeneracy between $M_{N_1}$ and $M_{N_2}$, where only NH allows 
the hierarchy $M_{N_1}\le M_{N_2}$, as shown in the bottom of Figs.~2. 
 \item
To explain the measured relic density without conflict of LFVs, 
$g_{3\mu}$ and $g_{3\tau}$ should be rather large, while  
$g_{3e}$ has to be small as shown in Figs.~3.
\item Once we satisfy the constraints of $\ell_\alpha \to \ell_\beta \gamma$
processes, the other current bounds on LFVs such as lepton universality 
and $\mu\mathchar`-e$ conversion are automatically satisfied in our framework.
\item The typical scale of the muon $g-2$ is $10^{-12}$ $\sim$  $10^{-11}$ 
with a negative sign, 
which has negligible effects on the deviation of the
experimental $g-2$ value of $O(10^{-9})$.
 \end{enumerate}
}

\appendix
\section*{ Appendix}
Here we explicitly show their formulas of Mandelstam valuables, and scalar products in terms of $v_{\rm rel}$ expanding form as follows:
\begin{align}
&s=(m_1+m_2)^2 +m_1m_2 v^2_{\rm rel},\\
&t=-\frac{m_1^2 m_2+m_1 (m_2^2-n_2^2)-m_2 n_1^2}{m_1+m_2}\nn\\
&+\frac{m_1m_2 v_{\rm rel} \cos\theta\sqrt{(m_1^2+2m_1m_2+m_2^2-n_1^2-n_2^2)^2-4n_1^2n_2^2}} {(m_1+m_2)^2}\nn\\
&-\frac{m_1m_2 v_{\rm rel}^2 \left(m_1^3+3 m_1^2 m_2+m_1(3m_2^2-n_1^2+n_2^2)+m_2(m_2^2+n_1^2-n_2^2)\right)} {2(m_1+m_2)^3}
,\\
&u=-\frac{m_1(m_1^2 + 2m_1m_2 +m_2^2-n_1^2+ n_1^2)}{m_1+m_2}\nn\\
&-\frac{m_1m_2 v_{\rm rel} \cos\theta\sqrt{(m_1^2+2m_1m_2+m_2^2-n_1^2-n_2^2)^2-4n_1^2n_2^2}} {(m_1+m_2)^2}\nn\\
&-\frac{m_1m_2 v_{\rm rel}^2 \left(m_1^3+3 m_1^2 m_2+m_1(3m_2^2+n_1^2-n_2^2)+m_2(m_2^2-n_1^2+n_2^2)\right)} {2(m_1+m_2)^3},
\end{align}
\begin{align}
&p_1\cdot p_2=\frac{s-m_1^2 - m_2^2}{2},\quad k_1\cdot k_2=\frac{s-n_1^2 - n_2^2}{2},\\
&p_1\cdot k_1=\frac{m_1(m_1^2+2m_1m_2+m_2^2+n_1^2-n_2^2)}{2(m_1+m_2)}\nn\\
&-\frac{m_1m_2 v_{\rm rel} \cos\theta\sqrt{(m_1^2+2m_1m_2+m_2^2-n_1^2-n_2^2)^2-4n_1^2n_2^2}} {2(m_1+m_2)^2}\nn\\
&+\frac{m_1m_2 v_{\rm rel}^2 \left(m_1^3+3 m_1^2 m_2+m_1(3m_2^2-n_1^2+n_2^2)+m_2(m_2^2+n_1^2-n_2^2)\right)} {4(m_1+m_2)^3}
,\\
&p_1\cdot k_2=\frac{m_1(m_1^2+2m_1m_2+m_2^2-n_1^2+n_2^2)}{2(m_1+m_2)}\nn\\
&+\frac{m_1m_2 v_{\rm rel} \cos\theta\sqrt{(m_1^2+2m_1m_2+m_2^2-n_1^2-n_2^2)^2-4n_1^2n_2^2}} {2(m_1+m_2)^2}\nn\\
&+\frac{m_1m_2 v_{\rm rel}^2 \left(m_1^3+3 m_1^2 m_2+m_1(3m_2^2+n_1^2-n_2^2)+m_2(m_2^2-n_1^2+n_2^2)\right)} {4(m_1+m_2)^3}
,\\
&p_2\cdot k_1=\frac{m_2(m_1^2+2m_1m_2+m_2^2+n_1^2-n_2^2)}{2(m_1+m_2)}\nn\\
&+\frac{m_1m_2 v_{\rm rel} \cos\theta\sqrt{(m_1^2+2m_1m_2+m_2^2-n_1^2-n_2^2)^2-4n_1^2n_2^2}} {2(m_1+m_2)^2}\nn\\
&+\frac{m_1m_2 v_{\rm rel}^2 \left(m_1^3+3 m_1^2 m_2+m_1(3m_2^2+n_1^2-n_2^2)+m_2(m_2^2-n_1^2+n_2^2)\right)} {4(m_1+m_2)^3}
,\\
&p_2\cdot k_2=\frac{m_2(m_1^2+2m_1m_2+m_2^2-n_1^2+n_2^2)}{2(m_1+m_2)}\nn\\
&-\frac{m_1m_2 v_{\rm rel} \cos\theta\sqrt{(m_1^2+2m_1m_2+m_2^2-n_1^2-n_2^2)^2-4n_1^2n_2^2}} {2(m_1+m_2)^2}\nn\\
&+\frac{m_1m_2 v_{\rm rel}^2 \left(m_1^3+3 m_1^2 m_2+m_1(3m_2^2-n_1^2+n_2^2)+m_2(m_2^2+n_1^2-n_2^2)\right)} {4(m_1+m_2)^3}
,
\end{align}
where $m_{1(2)}$ and $n_{1(2)}$ respectively represent the masses of initial state and final state.

\section*{Acknowledgment}  
This work was supported by the Ministry of Science and Technology
of Taiwan under Grants No. MOST-105-2112-M-007-028-MY3.

\end{document}